# Quantum entanglement of electrons in a biased 1D two-probe device

Yao-Sheng Li[1], Wen-Long You[1]*, and Xue-Feng Wang[1,2#]

[1]College of Physics, Optoelectronics and Energy, Soochow University, Suzhou, Jiangsu 215006, People's Republic of China

[2]Key Laboratory of Terahertz Solid-State Technology, Shanghai Institute of Microsystem and Information Technology, Chinese Academy of Sciences, 865 Changning Road, Shanghai 200050, China

* E-mail wlyou@suda.edu.cn
# E-mail: wxf@suda.edu.cn

## Abstract

Electronic quantum entanglement between the central chain and the two electrodes in an infinite one-dimensional two-probe device system is studied. The entanglement entropy is calculated employing the nonequilibrium Green's function method in the tight-binding model based on the relation between the correlation matrix and the von Neumann entropy. By extending the entropy to nonequilibrium cases, we have studied the scaling behavior when a voltage bias is applied between the two electrodes. The entropy usually decreases with the bias and may jump up when a quasi-state in the chain aligns in energy with the band edges in the electrodes. Odd-even effect is observed due to the symmetry of the chain.

**Introduction**

As a vital ingredient in quantum systems, entanglement is a useful resource for quantum information processing and quantum metrology [1]. Entanglement study in photonic systems has shown great application potential [2]. With the advance in nanotechnology, entanglement in electronic many-body systems is also attracting more and more attention and has been recognized as a precursor of quantum criticality [3] in condensed matter physics.

Among the proposed entanglement measures, von Neumann entanglement entropy is the most popular one. It has become a helpful bridge between several interdisciplinary theories, such as quantum gravity, quantum information and condensed matter physics. The entanglement entropy in the ground state of free fermions can be determined from the two-point correlation matrix of a fictitious Hamiltonian [4,5]. It works in any dimension, for arbitrary quadratic Hamiltonians, all states which are Slater determinants, and even at finite temperature. Thus it has been used in a large number of situations ranging from homogeneous chains to defect problems [6,7], random systems, higher dimensions, and the time evolution after a quench [8-11].

For a $d$-dimensional ($d$D) block composed by $N^d$ contiguous spins and embedded in an infinite many-body system, the von Neumann entropy of a *steady state* between the block and the rest of the system is given by [12,13]

$$S_N = -\text{Tr}\rho_N \log_2 \rho_N, \tag{1}$$

where $\rho_N$ is the reduced many-body density matrix for the $N^d$-site block. A celebrated boundary law of $S_N \sim N^{d-1} \propto$ surface area exists for the ground state of typical gapped and some gapless systems with short-range correlation. In gapless regimes, however, a logarithmic additive term of the form $S_N \sim \frac{c}{3} N^{d-1} \log_2 N$ dominates, which is explicit for 1D systems described by conformal field theories with a central charge $c$ [14-18]. In addition, the universal part of the entanglement spectrum, which is the eigenvalues $\xi_i$ of entanglement Hamiltonian $H_N$ descended from $\rho_N = e^{-H_N}$ [19] contains informations of intricate connection between bulk and edge properties [20].

Nanomaterials and nanodevices are practical assets to realize entanglement operations for quantum information technologies. Entanglement study on electronic systems with and without interaction or disorder may lead to quantum devices with cutting-edge functionalities. Electrons are usually far from equilibrium in the existence of electric bias or temperature gradient. However, straightforward computation of entanglement entropy from many-body density matrix over configurations costs too much and is currently restricted to equilibrium regime and small systems with few particles. Recently, non-equilibrium Green's functions (NEGF) method has been used to calculate correlation matrix and Neumann entropy for entanglement study for two-probe electronic systems [6].

In the last decades, NEGF method has been developed successfully in studying quantum transport in two-probe electronic devices [21]. It transfers many-body problems into single-particle ones and estimates current versus electric bias based on two-point Green's function and correlation matrix. In this paper, employing the NEGF method, we study the entanglement scaling of a one-dimensional (1D) two-probe tight-binding chain via the extended von Neumann entropy for non-equilibrium electrons.

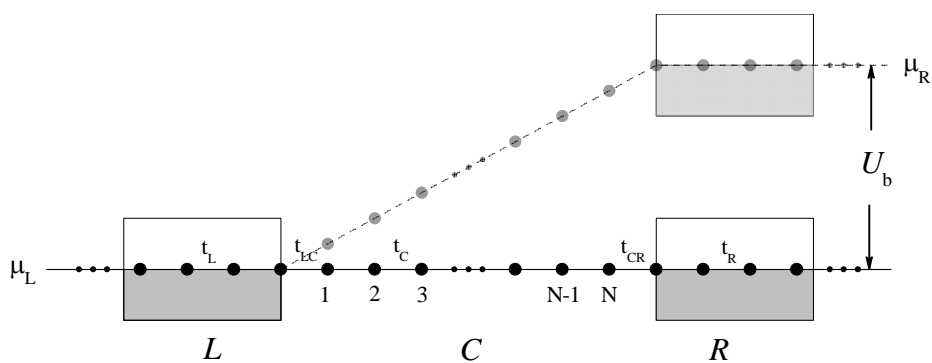

Fig. 1 Scheme of a two-probe device system of 1D chain composed of the electrodes $L$ and $R$ and the central part $C$. The potential profile of on-site

energies $\varepsilon_j$ under finite potential bias $U_b$ is illustrated by the upper chain.

**Model and method**

We consider a central chain $C$ of length $N$ sites attached to two infinite homogeneous electrodes $L$ and $R$. The system is described by the tight-binding Hamiltonian:

$$H = H_L + H_{LC} + H_C + H_{CR} + H_R, \qquad (2)$$

with $H_L = \sum_{j \leq 0}^{-\infty}(\varepsilon_j^L c_j^\dagger c_j + t_j^L c_{j-1}^\dagger c_j) + h.c.$ , $H_C = \sum_{j=1}^{N} \varepsilon_j^C c_j^\dagger c_j + \sum_{j=1}^{N-1} t_j^C c_j^\dagger c_{j+1} + h.c.$ ,

$H_R = \sum_{j \geq N+1}^{\infty}(\varepsilon_j^R c_j^\dagger c_j + t_j^R c_j^\dagger c_{j+1}) + h.c.$, $H_{CR} = t_{RC} c_N^\dagger c_{N+1} + h.c.$, and $H_{LC} = t_{LC} c_0^\dagger c_1 + h.c.$.

Here $c_j^\dagger$ ($c_j$) is the creation (annihilation) operator of electrons at site $j$, $t$ is the hopping integral between nearest neighbor sites, and $\varepsilon$ is the on-site energy. We assume a linear potential profile in the central chain when a bias voltage $U_b$ is applied between the two electrodes so $\varepsilon_j^L = 0$, $\varepsilon_j^R = U_b$, and $\varepsilon_j^C = jU_b/(N+1)$ as shown in Fig. 1.

The retarded $G^r$ and advanced $G^a$ Green's functions inside the central chain can be expressed in the energy space $\omega$ as [0b]

$$G^r(\omega) = [G^a]^\dagger = [EI - H - \Sigma_L - \Sigma_R + i\eta]^{-1} \qquad (3)$$

with the broadening functions $\Gamma_L = i(\Sigma_L - \Sigma_L^\dagger)$ and $\Gamma_R = i(\Sigma_R - \Sigma_R^\dagger)$, the self-energies $\Sigma_L = T_{LC}^\dagger G_L^s T_{LC}$ and $\Sigma_R = T_{RC}^\dagger G_R^s T_{RC}$, due to the coupling to electrode L and R, respectively. Here $G_L^s$ ($G_R^s$) is the surface Green's function of electrode L (R) which can be calculated self-consistently.

The corresponding correlation Green's function in the energy space $G^n(\omega) = G^r \Sigma^{in} G^a$ can then be obtained from the in-scattering function $\Sigma^{in} = \Sigma_L^{in} + \Sigma_R^{in}$ and the Fermi-Dirac distribution function $f(\omega)$ of temperature $T_e = 1/\beta k_B$ and Fermi energy $\mu$, where $\Sigma_L^{in} = \Gamma_L f_L(\omega) = \frac{\Gamma_L}{e^{\beta(\omega-\mu_L)}+1}$ and $\Sigma_R^{in} = \Gamma_R f_R(\omega) = \frac{\Gamma_R}{e^{\beta(\omega-\mu_R)}+1}$.

Integrating $G^n(\omega)$ in the whole energy space, we get the equal-time correlation function $G^n(t,t)$ with matrix element

$$G_{ij}^n(t,t) \equiv \langle c_j^\dagger(t) c_i(t) \rangle = \int_{-\infty}^{\infty} \frac{d\omega}{2\pi} G_{ij}^n(\omega) \tag{4}$$

and the diagonalized correlation matrix $\widehat{G}$ in its eigenvalue representation.

For free electrons in equilibrium, the reduced many body density matrix $\rho_N$ of order $O(2^N)$ can be derived from the correlation matrix of order $O(N)$ as [14]

$$\rho_N = \det(1-\widehat{G}) \exp\left\{ \sum_{i,j} [\ln \widehat{G}(1-\widehat{G})]_{ij} c_i^\dagger c_j \right\} \tag{5}$$

The entanglement entropy of the ground state is determined from the correlation matrix [12,13] and we have the expression of entanglement entropy

$$S = -Tr[\widehat{G} \log_2 \widehat{G} + (1-\widehat{G}) \log_2 (1-\widehat{G})]. \tag{6}$$

**Result and discussion**

In the following, we discuss the scaling of the entanglement defined by Eq. (6) in a homogeneous chain with $t_j^L = t_j^R = t_j^C = t_{LC} = t_{RL} = t = 1\,\text{eV}$ under a voltage bias $U_b$ between two half-filled electrodes with $\mu_X = \varepsilon_X$ for $X=L$ and $R$.

In Fig. 2, we plot the entanglement entropy $S$ for $N=2$ as a function of $U_b$. The entanglement entropy spectrum shows a non-monotonic behavior with peaks and valleys. The peak (valley) value decreases with the bias and a downward trend appears as a whole. The energy spectra of the integrand function $G_{ij}^n(\omega)$ and the in-scattering functions $\Sigma_L^{in}$ and $\Sigma_R^{in}$ are illustrated in Fig. 3 under biases $U_b = 0.1\,\text{V}$, 1 V, and the peak and valley values.

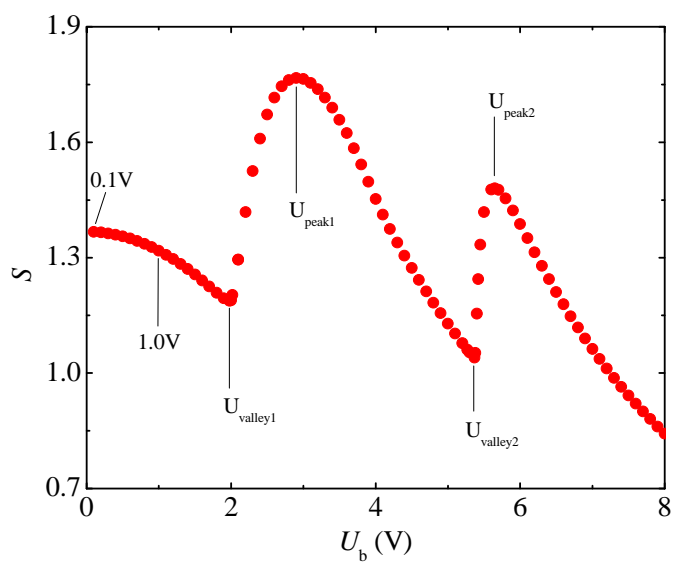

Fig.2. The entanglement entropy $S$ between a central chain C of size $N=2$ and the environment electrodes L and R as a function of bias $U_b$.

As shown in Fig. 3(a) and (g), for small bias $U_b = 0.1V$ when the system is in the linear regime, only the lower half of the energy band in the energy range $-2\text{eV} < \omega < 0$ is occupied by electrons. $\text{Re}[G_{ij}^n(\omega)]$ has a sharp peak at $\omega = -2$ eV due to the divergence of density of states (DOS) on the band edge and $\text{Im}[G_{ij}^n(\omega)]$ is negligible. $\text{Re}[\Sigma^{in}]$ and $\text{Im}[\Sigma^{in}]$, however, peak near $\omega = 0$ where the Fermi energy is located and strong scatterings occur. As $U_b$ increases, $\Sigma_R^{in}$ shifts $U_b$ while $\Sigma_L^{in}$ remain fixed. The peak of Re[$G_{ij}^n(\omega)$] shifts with $U_b$ too following the band edge of electrode R. The in-scattering function is finite only at energies in ranges $\mu_L - 2t < \omega < \mu_L$ or $\mu_R - 2t < \omega < \mu_R + 2t$ where electrons exist in the electrodes. The diagonal elements of the imaginary part Im[$G_{ii}^n(\omega)$]=0 and the nondiagonal elements Im[$G_{12}^n(\omega)$] and Im[$G_{21}^n(\omega)$] have finite value with opposite sign in energy range $Max\{\mu_L, \mu_R - 2t\} < \omega < Min\{\mu_R, \mu_L + 2t\}$. The entropy S decreases with $U_b$ when it is small.

When the Fermi energy $\mu_R$ of electrode R touches the band edge of electrode L at $U_b = 2$ V, another sharp peak appears for the diagonal elements Re[$G_{ii}^n(\omega)$] and a negative peak appears for non-diagonal elements Re[$G_{12}^n(\omega)$] at energy $\omega = \mu_R = 2$ eV as shown in Fig. 3(c). The entropy increases quickly with the emerging of the new positive and negative peaks and shows a valley at $\omega = 2$ eV. The previous low energy peak of Re[$G_{22}^n(\omega)$] near the band edge of electrode R at energy $\omega = eU_b - 2t$ begins to decay gradually with $U_b$. The entropy reaches a peak as it disappears at $\omega = 2.9$ eV and then decreases slowly as shown in Fig. 2. These positive and negative peaks of Re[$G_{ij}^n(\omega)$] shift to lower energy when $U_b$ continues increasing and reach to the band edge of electrode R at $U_b = 5.37$ V as shown in Fig. 3(e). At this bias, the entropy jumps to higher values again as shown in Fig. 2. Under higher $U_b$, the Re[$G_{ij}^n(\omega)$] peaks migrate to higher energy again and Re[$G_{ij}^n(\omega)$] becomes negligible outside the energy range $\mu_R - 2t < \omega < \mu_R$.

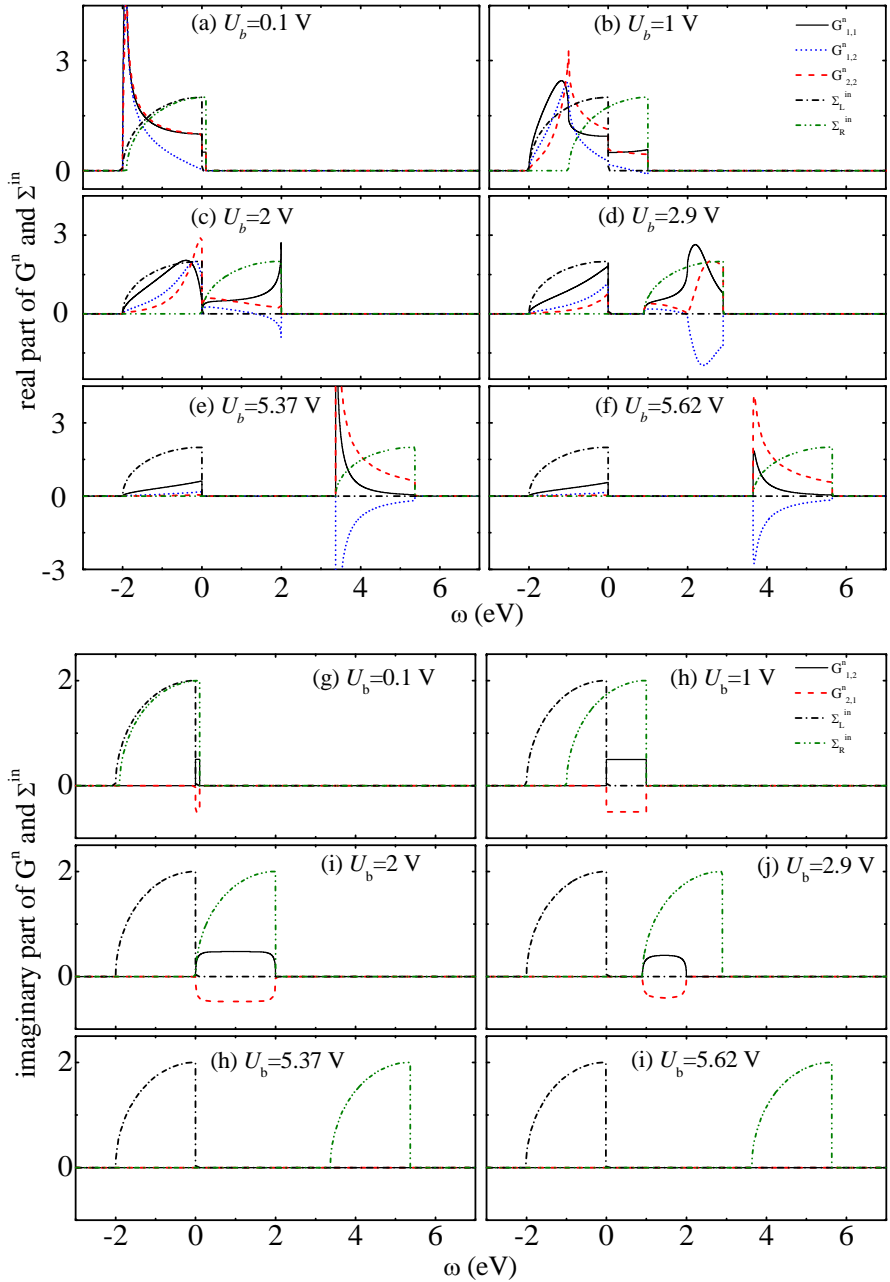

Fig.3. The real and imaginary energy spectrum of the correlation function elements $G_{ij}^n(\omega)$ and $\Sigma^{in}$ at biases $U_b$ =0.1, 1, 2, 2.9, 5.38, 5.62 V for $N$=2 and $t$=1 eV. $\mathrm{Re}[G_{21}^n(\omega)] = \mathrm{Re}[G_{12}^n(\omega)]$ and $\mathrm{Im}[G_{11}^n(\omega)] = \mathrm{Im}[G_{22}^n(\omega)] = 0$ are not shown.

The elements of correlation matrix are list in Table 1 for bias values when the entropy reaches valley and peak. The sum of the eigen values $\hat{G}_1$ and $\hat{G}_2$ remains unit for $U_b<5.37$ V, corresponding to a half filled system. Under a bias of value above 5.37V, the central chain begins to lose electrons quickly and becomes less filled. The occupation rate reaches 31% at $U_b=5.62$ V.

Table 1. The matrix elements of the equal time correlation function $G^n(t,t)$, its eigenvalues $\hat{G}_1$ and $\hat{G}_2$, and the entanglement entropy S at biases $U_b$ =0.1, 1, 2, 2.9, 5.37, and 5.62 V for $N=2$.

| $U_b$ (V) | 0.1 | 1 | 2 | 2.9 | 5.37 | 5.62 |
|---|---|---|---|---|---|---|
| $\text{Re}[G^n_{1,1}(t,t)]$ | 0.502 | 0.525 | 0.624 | 0.676 | 0.354 | 0.205 |
| $\text{Re}[G^n_{1,2}(t,t)]$ | 0.318 | 0.319 | 0.301 | −0.064 | −0.354 | −0.179 |
| $\text{Re}[G^n_{2,2}(t,t)]$ | 0.498 | 0.475 | 0.376 | 0.324 | 0.645 | 0.422 |
| $\text{Im}[G^n_{1,2}(t,t)]$ | $7.96E-3$ | $7.90E-2$ | 0.144 | 0.063 | $2.4E-9$ | $9.8E-10$ |
| $\hat{G}_1$ | 0.182 | 0.171 | 0.144 | 0.302 | 0.116 | 0.104 |
| $\hat{G}_2$ | 0.818 | 0.829 | 0.856 | 0.698 | 0.883 | 0.523 |
| S | 1.367 | 1.318 | 1.189 | 1.767 | 1.040 | 1.481 |

To understand the jump behavior of entanglement entropy and the out-flowing of charge in the central chain under voltage bias, we calculate the spectral function from the Green's function $G^r$ given in Eq.(3). The main peaks of the spectral function correspond to the quasi eigenstates of the effective Hamiltonian defined by $G^r$. In Fig. 4 we plot the entropy, electron occupation of energy bands in the electrodes, and the energy of the quasi states as functions of the bias for $N=1$-4. It is shown that the sudden turning and the oscillation of the entropy versus bias curve correspond to the crosses of quasi state energy in the central chain with the band edges in the electrodes.

For $N=1$ there is mainly one quasi state as shown by the blue curve in Fig.4(a2) . As the bias increases the energy band of electrode R (in olive) goes up. Under $U_b<4$ V, the energy bands of the electrodes overlap and the quasi state remains inside the bands. The one-site central chain is half occupied and the entropy remains constant. When $U_b>4$ V, the electrode bands separate with each other without overlapping and the quasi state goes out of the bands. The electron number inside the chain decreases and the entropy decreases too. In the $N=2$ case, the entropy decreases with the bias under low bias. When the occupied energy regimes of the two bands separate from each other at $U_b=2$ V, one quasi state go out of the bands but follows closely with the band-R edge until $U_b=2.9$ V. Correspondingly the entropy increases and then decreases with the bias showing a valley at $U_b=2$ V and a peak at $U_b=2.9$ V. Near $U_b=5.4$ eV this quasi state touches and then leave from the upper band-L edge. In the same energy range, another quasi state emerges out from the band-R bottom and results in another pair of entropy valley and peak at $U_b=5.37$ and 5.62 V, respectively. Since the two states are now out of the energy range of the electrode bands, they become bound states and less occupied. As a result, the total occupancy in the central chain given by $\hat{G}_1+\hat{G}_2$ begins to decrease at $U_b=5.37$ eV as indicated in Table 1. The above result suggests that when the quasi state in the central chain degenerates with the electrode states near energy bands, the entropy remains relatively in a high level due to high DOS on band edges and strong coupling between the central chain and the electrode. When the quasi states leave from the energy regime of the bands, they decouple from the electrodes to become less occupied bound states and the entropy decreases.

Similar behavior can be observed in systems with central chain of different length. $N=4$. As shown in Fig. 4(c1) and (c2) for $N=3$, the entropy begins to increase when one quasi state near the upper band-L edge enters the occupied energy range of band-R at $U_b$=2 V. When the quasi state leave from the bands and become a bound state at $U_b$=4 V, the entropy decrease quickly until other quasi states appear near the band edges at $U_b$=7.1 V. Similar to the case $N=1$, at $U_b$=4 V, the overlapping between the two electrode bands vanishes and a quasi-state in the chain 'leave' from the edge cross of the two bands, a derivative discontinuity occurs on the $S$-$U_b$ curve. For $N=4$, similar to the case of $N=3$, the approach of a quasi-state to the lower edge of band-R lead to a turning of the $S$-$U_b$ curve from decrease to increase at $U_b$=2 V. The quasi-state then deviates slowly from the band edge and is well away from the band edges when they separate at $U_b$=4 V. As a result, different from the cases for odd $N=1$ and 3, there is no derivative discontinuity in the $S$-$U_b$ curve for even $N=2$ and 4. Around $U_b$=4.7 V, another quasi state emerge from the outer side of band-R and an entropy valley appears. The entropy increases and then decreases when the quasi-state is away from the band edge and becomes a well-defined bound states.

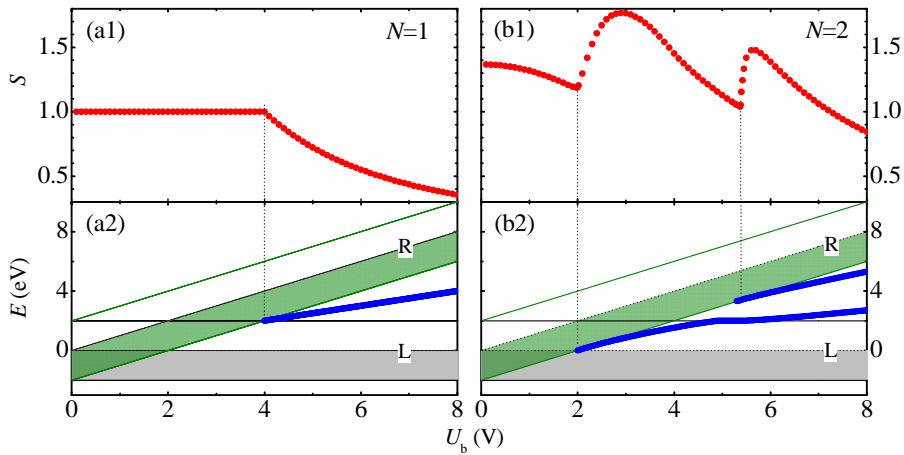

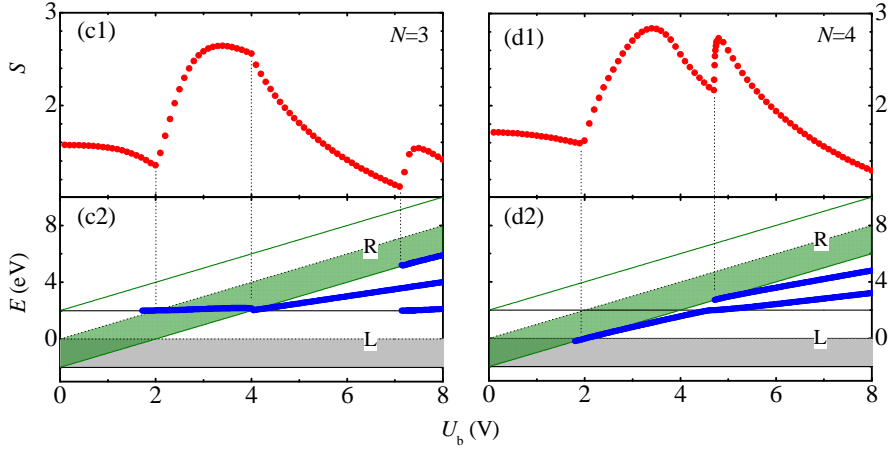

Fig.4. The entanglement entropy (red circles) is plotted function of the voltage bias $U_b$ for chain length $N=1$- $4$ in (a1)-(d1), respectively. The corresponding energy bands of electrodes and the energies of quasi-states (blue dots) of the effective chain Hamiltonian defined by $G^r$ are shown in (a2)-(d2). The black (olive) solid frame illustrates band-L (R) of electrode L (R) with grey (olive) shadow for occupied states.

The scaling of the entanglement entropy is shown in Fig. 5 for different voltage bias. In a system without bias, $U_b=0$, the system is a trivial 1D chain of free electrons and we have $S_L \sim \frac{1}{3}\log_2 N$ as expected. The scaling behavior of the system follows well the law under low bias but deviates clearly when $U_b > 1$ V especially for short central chain. As the bias increases, the two electrode energy bands separate from each other and the quasi-state energy inside the central chain crosses with the band edges. The degeneracy of the quasi-state and the band edge states results in a boost of entanglement between the central chain and the electrodes. The scaling curve become oscillating instead of monotonic.

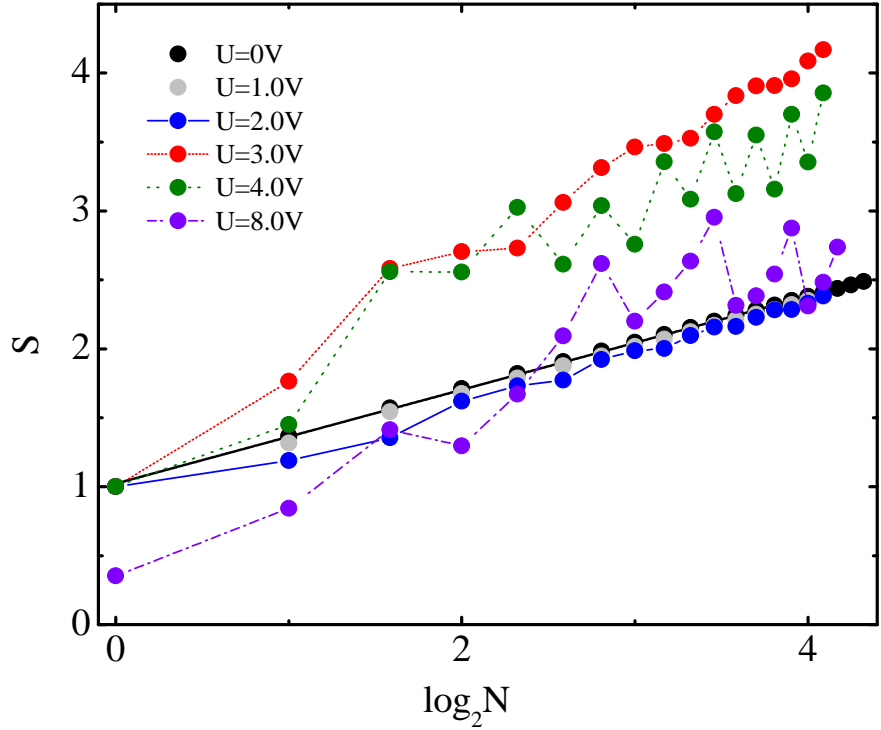

Fig.4. The entanglement entropy between the central chain and the electrodes is plotted as a function of $\log_2 N$ for voltage bias $U_b$ =0, 1, 2, 3, 4, and 8 V. The thick solid line shows the linear fitting of the zero bias result which obeys the well-known scaling law $S \sim \frac{1}{3}\log_2 N$.

**Conclusion**

Employing the nonequilibrium Green's function theory in the tight-binding model, we have studied the entanglement between a one-dimensional chain of finite size and two attached infinite electrodes in nonequilibrium states. Voltage bias between the electrodes reduces the entanglement entropy in the linear regime and can enhance the entanglement when quasi-states coincide in energy with the edge of electrode energy bands. This results in oscillation of the entropy as the quasi states align with the band edge one by one with the bias increase. The entropy has an over whole downward trend indicated by the energy decrease of entropy peaks and valleys with the increase of bias. This behavior is impressively different from that of electron transport between the electrodes which disappears when the overlapping between the energy bands of the electrodes vanishes under high bias. Even-odd effect is observed in the scaling of the entanglement under finite bias.